# Application of the Monte Carlo Method in Modeling Transport and Acceleration of Solar Energetic Particles

Valeriy Tenishev[1], Lulu Zhao[1], Igor Sokolov[1]

[1]University of Michigan

**Key Points:**

- Kinetic model AMPS coupled with SC, IH, and OH components of the SWMF solves the focus transport and Parker equation to simulated the dynamics of SEPs in the heliosphere
- Modeling SEPs beyond 1 AU is important for accurate prediction of a SEP event decay phase
- The choice of FTE vs Parker equations are important outside of the region of the particle acceleration by a shock

Corresponding author: Valeriy Tenishev, `vtenisge@umich.edu`

# 1 Introduction

The need for quantitative characterization of the solar energetic particle (SEP) dynamics goes beyond being an academic discipline only. It has numerous practical implications related to human activity in space. The terrestrial magnetic field shields the International Space Station (ISS) and most uncrewed missions from exposure to SEP radiation. However, extreme SEP events with hard energy spectra are particularly rich in hundreds of MeV to several GeV protons that can reach the altitudes of the Low Earth Orbit (LEO). These protons have a high penetrating capability, thus producing significant radiation hazards for human spaceflight (Jäkel, 2004; Hellweg & Baumstark-Khan, 2007). Outside of the geospace protection by the geomagnetic field, the radiation risks multiply. The main radiation hazard to humans is mainly due to protons with energy > 30 MeV (Mewaldt, 2006).

SEPs also have a significant effect on the atmosphere. Sudden ionization of the upper atmosphere at high latitudes that occurs during polar cap absorption (PCA) events can block high frequency (HF) communication for hours, affecting communication with aircraft on the intercontinental high altitude flights (Mewaldt, 2006; Morris, 2007). Another effect of SEPs in the atmosphere is creating $NO_x$ molecules in the upper atmosphere that can deplete the atmospheric ozone population. Such, during the October 28–31, 2003 period, ozone depletion of 5% to 8% was measured in the southern polar stratosphere that lasted days beyond the events (Jackman et al., 2005).

Here we present the first results of the application of the Adaptive Mesh Particle Simulator (Tenishev et al., 2021) for modeling the transport and acceleration of solar energetic particles in the heliosphere. The paper also presents an analysis of (1) how various pitch angle diffusion coefficient approximations affect the properties of the simulated SEPs population and (2) discusses how pitch angle scattering when SEPs are beyond 1 AU affects a SEP event decay phase at the Earth's orbit.

# 2 Source, acceleration and transport of solar energetic particles (SEPs) in the inner heliosphere

SEP events are generally classified as impulsive, and gradual (e.g., Li et al., 2012; Qin & Wang, 2015). Solar flares produce impulsive events. They are short and characterized by low intensity, enhanced electrons and $^3$He density, and high charge states of heavy ions for a broad energy range. In contrast, gradual events have high protons flux, last for days, and are characterized by relatively low charge states of heavy ions (Zhang et al., 2009; Qin & Wang, 2015; Hu et al., 2017). The sources of the high energy SEPs in gradual events is located close to the solar corona (Qin et al., 2011, 2013; He et al., 2011; Zhang et al., 2009). Lower-energy SEPs can be produced at a CME shock up to distances of several AU (Cane et al., 1988; Reames, 1999).

The differential intensity, $j(E) = j_0 E^{-\gamma} exp(-E/E_0)$, where $\gamma$ is a normalization factor, and $E_0$ is a function of rigidity of the accelerated species, has been successful in representing the energy spectra of protons, electrons, and $\alpha$-particles during SEP events (Mewaldt, 2006; Cohen et al., 2003; Desai et al., 2004). Band et al. (1993) has suggested a more complex double power-law differential intensity model. Detailed surveys of CME-driven shocks, associated SEP events, and ion composition and spectra during SEP events were presented by e.g., Desai et al. (2003), Ho et al. (2003) and Cohen et al. (2007).

Observations of SEP events by Cane et al. (2003) demonstrate that the time-intensity profiles exhibit two peaks, with an earlier one having a high Fe/O ratio and the subsequent one having a low Fe/O ratio. A possible explanation of this phenomenon is that these two-component events are due to a flare and a subsequent CME. In such a model, the first peak is caused by the flare, and the second peak is due to particle acceleration at the CME-driven shock (e.g., Li & Zank, 2005; Li et al., 2012).

Simultaneous spacecraft observations of a SEP event (e.g., by Helios 1 and 2, or at different latitudes and radial distances, or by ACE and Ulysses) reveal unexpected phenomena. Depending on their location, observing spacecraft are magnetically connected to different parts of a CME shock. At the beginning of an event, such observations demonstrate significant difference in the energetic particle intensity that would depend on the magnetic connectivity to the source of SEPs (e.g., Qin et al., 2013). During the decay phases, the spatial gradients of SEP fluxes are diminished in all latitudinal, longitudinal, or radial directions, and the SEP fluxes measured by widely separated spacecraft typically present similar intensities within a small $\sim$ 2–3 factor (Mckibben, 1972; McKibben et al., 2003; Qin & Wang, 2015; Zhang et al., 2009). The phenomenon is called the "reservoir phenomenon" (Roelof et al., 1992). The reservoir phenomena is not specific to protons only. Such, Lario et al. (2013) report that similar electron intensities at MESSENGER and near 1 AU spacecraft have been observed in the decay phase of several events (e.g., Lario & Decker, 2011). After reaching a uniform reservoir the particle intensity everywhere decreases at an approximately the same rate in time, which can sometimes last for more than a month (Zhang et al., 2009; Reames et al., 2013). There is no definite understanding the nature of the reservoir phenomena. Two possible mechanisms are suggested. Mckibben (1972) and McKibben et al. (2003) suggested that perpendicular diffusion distribute the particles uniformly. Another direction of thought is that an enhanced magnetic field and turbulence shortly behind shock mirror particles back preventing them from propagating into the outer heliosphere (Reames et al., 2013; Qin et al., 2013; Tan et al., 2009; Roelof et al., 1992).

## 2.1 Transport and Acceleration of SEPs in the heliosphere

It is commonly accepted that modeling SEPs in the heliosphere can be done assuming particle propagation along meandering magnetic field lines connecting the source of SEPs and the observer. Laitinen et al. (2013) suggest that particle diffusion across the field lines is negligible during the first 5 hours from the beginning of an event. Accounting for the fact that (1) characteristic scales of the simulated phenomena (transport of SEPs in the inner heliosphere) significantly exceed those related to the gyro-motion of the energetic particles, and (2) the drift velocity that is associated with averaging of gyration of the energetic particles is small, the assumption of SEPs transport along evolving magnetic field lines has a solid physics reasoning.

A particle moving along a magnetic field line experiences adiabatic focusing, adiabatic cooling, pitch angle scattering, and stochastic acceleration (e.g., Kota et al., 2005; Hu et al., 2017; Schwadron et al., 2010). The combination of those processes is accounted for in the focused transport equation that describes the transport of energetic particles along a magnetic field line and temporal evolution of the particle momentum and pitch angle (Skilling, 1971; Ruffolo, 1995; Ng & Reames, 2003; Tylka, 2001; He et al., 2011; He & Wan, 2019; Qin et al., 2006; Zhang et al., 2009; Li et al., 2003)

$$\frac{\partial f}{\partial t} + \mu v \frac{\partial f}{\partial z} + \mathbf{V}^{sw} \cdot \nabla f + \frac{dp}{dt}\frac{\partial f}{\partial p} + \frac{d\mu}{dt}\frac{\partial f}{\partial \mu} - \frac{\partial}{\partial \mu}\left(D_{\mu\mu}\frac{\partial f}{\partial \mu}\right) = Q(\mathbf{x}, \mathbf{p}, t), \quad (1)$$

where $f$ is the distribution function, $\mu$ is a pitch angle, $v$ is a particle velocity, $p$ is a particle momentum. The effect of stochastic forces on the particle's pitch angle is described with the diffusion coefficient $D_{\mu\mu}$. Finite volumes, finite difference, and Monte Carlo methods have been successfully used for solving Eq. 1. Here we will use Adaptive Mesh Par-ticle Simulator (AMPS) for solving Eq. 1 (Tenishev et al., 2021). In the employed Monte Carlo method, the population of the simulated SEPs is represented by a large number of the model particle, which is affected by the same processes and forces as actual ions in the simulated SEPs populations. Hence, the particles' momentum and pitch angle are incremented during a simulation time step as follows (e.g., He et al., 2011):

$$dz(t) = \mu v dt \quad (2)$$

$$dp = -p \left[ \frac{1-\mu^2}{2} \left( \frac{\partial V_x^{sw}}{\partial x} + \frac{\partial V_y^{sw}}{\partial y} \right) + \mu^2 \frac{\partial V_z^{sw}}{\partial z} \right] dt \quad (3)$$

$$d\mu = \left[ \frac{1-\mu^2}{2} \left[ \frac{v}{L} + \mu \left( \frac{\partial V_x^{sw}}{\partial x} + \frac{\partial V_y^{sw}}{\partial y} - 2\frac{\partial V_z^{sw}}{\partial z} \right) \right] \right] dt + \sqrt{2D_{\mu\mu}} dW_\mu(dt), \quad (4)$$

where $z$ is the distance along the magnetic field line, $\mu = \cos\theta$ is the particle pitch angle cosine, and t is the time. The macroscopic forces are characterized by $L(z) = B(z)/(-\partial B/\partial z)$, the focusing length in the diverging magnetic field. $dW_\mu(dt) = dt \cdot \Lambda$ denotes a Wiener process, where $\Lambda$ is a Gaussian distributed random number. Some of the recent examples of SEPs models based on solving the focused transport equation (Eq. 1) were presented by e.g., Li et al. (2003), Hu et al. (2017), Schwadron et al. (2010), He and Wan (2015), Dresing et al. (2012), Zhang et al. (2009), Dröge et al. (2010, 2006).

### 2.2 Effect of the pitch-angle scattering

Table 1 summarizes the formulation of the pitch angle diffusion coefficient that we have found in the literature. These diffusion coefficients are implemented in AMPS and available for the CCMC users to model the propagation of SEPs in the inner heliosphere. In the future, we plan to enhance the capabilities of AMPS available in the CCMC to study the transport of SEPs and galactic cosmic rays (GCRs) in full 3D.

Below is the description of the parameters of individual diffusion coefficient models summarized in Table 1. The effect of the diffusion coefficient choice on the result of the modeling transport of SEPs is discussed later in the Summary Section.

Type I (Borovikov et al., 2019): $r_{L0}$ = 1 GeV/$ceB$ is a Larmor radius for the particle momentum 1 GeV/c. $L_{max}$ is the maximum spatial scale in the turbulence, which is related to the minimum wave vector as $Lmax = 2\pi k_L^{-1}$.

Type II (le Roux & Webb, 2009): $V_A$ is the Alfvén speed, $\epsilon$ is the ratio of the magnetic energy density of backward propagating Alfvén waves divided by the magnetic energy density of forward propagating Alfvén waves, $\Omega$ is the gyrofrequency of the energetic charged particles, $l_b$ is the wavelength at which there occurs a break in the power spectrum of magnetic field fluctuations separating the energy range from the inertial range, and

$$D_0 = \frac{\pi}{8} A^2 \Omega^2 l_b \left(1 - \mu^2\right) \quad (5)$$

Here $A^2 = \langle \delta B^2 \rangle / B^2$ is the average total energy density of magnetic field fluctuations associated with Alfvén waves normalized to the energy density in the background magnetic field, $V_A \sim 38$ km s$^{-1}$. $l_b = 0.03 \ \bar{r}$ AU upstream, based on observations of the radial dependence of the MHD turbulence correlation scale in the solar wind, kept $l_b$ constant across the termination shock, and let it decrease as $l_b \sim r^{-1}$ downstream, using the MHD turbulence transport theory of Zank et al. (1996) as a guide; assumed simply that $\epsilon = 1$ (equal amount of forward and backward propagating Alfvén waves) everywhere; and specified upstream that $\langle \delta B_\perp^2 \rangle = 0.1 B_0^2 (1\,\text{AU}/r)^{3.5}$. Hu et al. (2017) suggests the turbulence level of $\delta B(r)^2 \sim \delta B(1\,\text{au})^2 \frac{1\,\text{au}}{r}$ with $\delta B^2/B^2 = 0.1$ at 1 au, and the turbulence power spectrum $P^{slab}(k) = A_s \lambda_c \delta B_{slab}^2 \left[1 + (k\lambda_c)^2\right]^{-\sigma/2}$, where $\sigma = 5/3$, correlation length $\lambda_c = 10^9$ m (Zank et al., 2004), $A_s$ is such that $\int_{k_R} P^{slab}(k) dk = \delta B_{slab}^2$, and $k = \gamma m \Omega / p$.

Table 1. Summary of the pitch angle diffusion coefficient approximations

| Type | References | Formulation |
|---|---|---|
| I | Borovikov et al. (2019) | $D_{\mu\mu} = \dfrac{v}{\lambda_{\mu\mu}} (1-\mu^2) \|\mu\|^{2/3}$ <br><br> $\lambda_{\mu\mu} = 0.5 \dfrac{B^2/\mu_0}{w_- + w_+} (L_{\max}^2 r_{L0})^{1/3} \left(\dfrac{pc}{1\,\text{GeV}}\right)^{1/3}$ |
| II | le Roux and Webb (2009) | $D_{\mu\mu} = D_0 \dfrac{1}{1+\epsilon}\left(1 - \dfrac{\mu V_A}{v}\right)^2 \dfrac{\|v\mu - V_A\|^{2/3}}{\|v\mu - V_A\|^{5/3} + (\Omega l_b)^{5/3}} +$ <br> $D_0 \dfrac{\epsilon}{1+\epsilon}\left(1 + \dfrac{\mu V_A}{v}\right)^2 \dfrac{\|v\mu + V_A\|^{2/3}}{\|v\mu + V_A\|^{5/3} + (\Omega l_b)^{5/3}}$ <br><br> $D_0 = \dfrac{\pi}{8} A^2 \Omega^2 l_b (1-\mu^2),$ |
| III | Hu et al. (2017); Jokipii (1966); Zhao and Li (2014) | $D_{\mu\mu} = \dfrac{\pi}{4}(1-\mu^2)\Omega_0 \dfrac{k P^{slab}(k)}{B^2}$ <br><br> $P(k) = A_\beta \dfrac{\lambda_c (\delta B)^2}{1 + (k\lambda_c)^\beta}$ |
| IV | Wang and Qin (2004) | $D_{\mu\mu} = D_0 v p^{q-2}(\|\mu\|^{q-1} + h)(1-\mu^2)$ |
| V | Qin et al. (2013) | $D_{\mu\mu} = \left(\dfrac{\delta B_{slab}}{B_0}\right)^2 \dfrac{\pi(s-1)}{4s} k_{min} v R^{s-2}(\mu^{s-1} + h)(1-\mu^2)$ |

Type III (Zhao & Li, 2014): Turbulence power spectrum is in the form

$$P(k) = A_\beta \frac{\lambda_c (\delta B)^2}{1 + (k\lambda_c)^\beta}, \quad (6)$$

where $\beta = 5/3$, and $A_\beta$ is a normalization constant defined as

$$\int_0^\infty P(k) dk = (\delta B)^2 \quad (7)$$

Here, $\lambda_c = 10^9$ m is a correlation length. The typical values at 1 au: $k_L = 2.0 \times 10^{-7}$ m$^{-1}$, $k_R = 1.0 \times 10^{-10}$ m$^{-1}$, $\lambda_c = 10^9$ m, and $(\delta B/B_0)^2 = 0.05$.

Type IV (Wang & Qin, 2004): Here, $D_0$ is a constraint that controls the magnetic field fluctuations level. The constant $q$ is chosen 5/3 for a Kolmogorov spectrum type of the power spectral density of magnetic field turbulence in the inertial range. Furthermore, h = 0.01 is chosen for non-linear effect of pitch angle diffusion at $\mu = 0$ in the solar wind (Qin & Shalchi, 2009).

Type V (Qin et al., 2013): The diffusion coefficient parameters are $\delta B_{slab}$ is the magnitude of the slab component of the turbulence, $s$ is the spectral index in the inertial ranges, $k_{min}$ is the lower limit of the wave number of the inertial range in the slab turbulence power spectrum, $R = pc/(qB_0)$ is the particle Larmor radius, and $q$ is the particle charge. $h$ comes from the nonlinear effect of magnetic turbulence on the pitch-angle diffusion at $= 0$ (Beeck and Wibberenz 1986; Qin & Shalchi 2009). In our simulations, we set $s = 5/3$, $k_{min} = 1/l_{slab} = 33 AU^{-1}$, and $h = 0.01$, where lslab is the slab turbulence correlation length. $\delta B_{slab}$ is assumed to be proportional to the magnitude of the local background magnetic field.

### 2.3 Seed population

Suprathermal particles are more easily passing the injection threshold (~ 100 keV/nuc) than are thermal ions from the solar wind (e.g., Tylka et al., 2006; Zhao & Li, 2014). Hence, these pre-accelerated particles are more easily accelerated to high energy at the front of a CME-driven shock. There are two primary sources of suprathermal particles that are available for further acceleration in a CME-driven shock wave. Thermal solar wind particles can be heated by a solar flare that is preceded by an ICME shock. Those suprathermal particles form a source for particles capable of being injected into a shock for further acceleration. This mechanism produces a two-peak SEP event discussed in Section 2. Another source of suprathermal particles is the reservoir (see Section 2), which contains particles left behind a CME-driven shock (Cliver, 2006; Reames et al., 2013). This conclusion is supported by observations. Such, Desai et al. (2003, 2004) suggested that a seed population predominantly comprised ions that were previously accelerated in impulsive and gradual SEPs and that the shock acceleration process accelerated higher rigidity ions less efficiently than lower rigidity ions.

Usually, a power law energy spectrum of seed population particles is assumed

$$f(E) = f(E_0) (E/E_0)^{-\delta}, \quad (8)$$

where $E_0 = m_p U_{up}^2 /2$ is the injection energy of a proton in a parallel shock, and power law spectra index of $\delta \geq 3.5$ (Desai, Mason, Gold, et al., 2006; Desai, Mason, Mazur,

& Dwyer, 2006; Hu et al., 2017). In more detail, injection and diffusion coefficient in the case of oblique waves is discussed by Li et al. (2012).

The power-law distribution of the seed population particle energy (Eq. 8) was used in the further modeling work.

(Young et al., 2021): A more complex model of the seed population

$$f(E,r) = \frac{m_p^2}{2E} \frac{J_0}{\xi} \left(\frac{r_1}{r}\right)^\beta \left(\frac{E}{E_r}\right)^{-\gamma} \exp\left(-\frac{E}{E_0}\right), \qquad (9)$$

where m is the particle mass, $J_0$ is the amplitude of the 4He fluence spectrum, $\xi$ is the ratio of 4He to H, $r_1$ is a fixed reference distance, $\beta$ is the radial scaling dependence, $E_r$ is a reference energy, $\gamma$ is the energy power-law dependence, and $E_0$ is the rollover energy.

# 3 Methodology

Extreme SEPs are usually associated with CME eruptions. The connection is deep: the interaction of the outward propagating CME with the ambient solar wind produces the interplanetary shock wave, which is generally thought of as the main factor in the acceleration of SEPs.

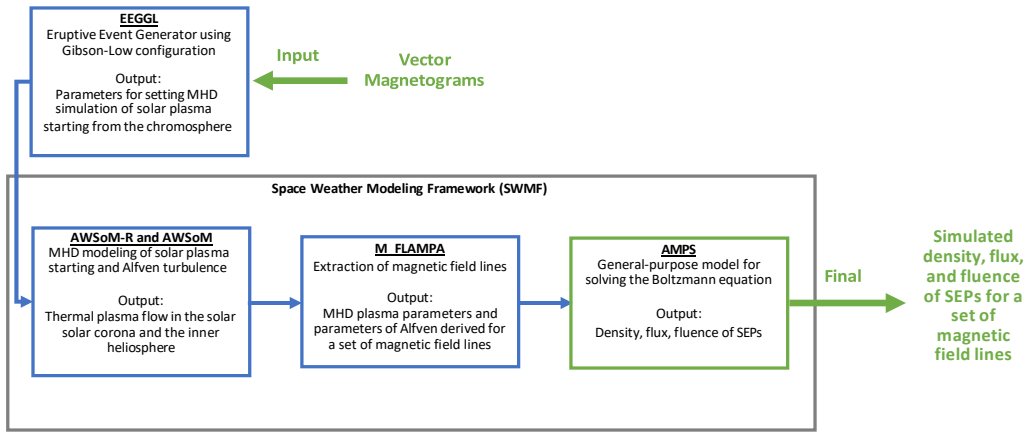

**Figure 1.** The figure illustrates the implemented coupling approach. First, we use EEGL is used to locate the place of a flare eruption and to derive the initial conditions for further modeling of solar plasma.

## 3.1 Eruptive Event Generator

EEGGL (Eruptive Event Generator using Gibson-Low configuration) is a user-friendly tool developed by Jin et al. (2017, 2017) at the U-M and transitioned to the CCMC at NASA GSFC. It integrates solar images of the eruption into an intuitive visual user interface that allows the user to set the parameters of the GL magnetic configuration, which is designed to model a magnetic-driven CME and its propagation to 1 AU. EEGGL incorporates magnetograms of the solar magnetic field prior to the eruption, and, if possible, the multi-point observations of the CME near the Sun. The CME speed is obtained with the help of the STEREOCat web-application available at the CCMC, which allows the user to derive both the CME speed and an approximate source location. With these input parameters, the operator chooses the active region (AR) from which the CME originates. Using these inputs EEGGL automatically (i) processes the magnetogram; (2) analyzes and calculates the integral parameters of the AR; (3) automatically sets the parameters of the GL magnetic configuration; and finally (4) visualizes the magnetic field of the AR and the GL configuration to verify that they match.

### 3.2 Global MHD plasma simulation from the upper chromosphere and to 1 AU

This section illustrate the modeling approach used for calculating the background solar wind starting the upped chromosphere and to 1 au that is used in further modeling transport and acceleration of SEPs.

#### 3.2.1 Block-Adaptive-Tree Solar-wind Roe-type Upwind Scheme (BATS-R-US) Code

The BATS-R-US (Block-Adaptive-Tree Solar-wind Roe-type Upwind Scheme) code was developed at the U-M. It solves the equations of extended MHD – a system of equations describing the transport of mass, momentum, energy, and magnetic flux (Groth et al., 2000; Powell et al., 1999). This high-performance code enables Sun-to-Earth CME propagation simulations to be performed in near real-time when running on hundreds of processors on a supercomputer (Manchester, Gombosi, Roussev, Ridley, et al., 2004). The implementation of adaptive-mesh-refinement (AMR) in BATS-R-US allows orders of magnitude variation in numerical resolution within the computational domain while keeping the total computational resource requirement at a reasonable level. That is important for a global model of the magnetized solar plasma in which one strives to resolve small structures like shocks, flux ropes, electric current sheets in a 3-D domain, which may extend to hundreds of $R_\odot$. In the context of solar-heliospheric physics, BATS-R-US has been utilized to model the global structure of the solar corona and solar wind, the initiation (Roussev et al., 2003; Jacobs et al., 2009) and propagation of idealized (Manchester, Gombosi, Roussev, De Zeeuw, et al., 2004; Manchester, Gombosi, Roussev, Ridley, et al., 2004) and more realistic (Roussev et al., 2004, 2007; Evans et al., 2011; Jin et al., 2013) solar eruptions and associated SEP events (Sokolov et al., 2004, 2009).

In the work presented in this paper, in the SC, IH regions, we have used the BATS-R-US-based AWSoM model that self-consistently taking into account wave turbulence. In the SC, the AWSoM-R model (see Section 3.2.2) is applied above the inner boundary at the heliocentric distance of $R = 1.1\text{- }1.15\ R_\odot$, while below the boundary the model is bridged to the upper chromosphere via threaded field lines. This way, we save computational resources that otherwise would be spent to resolve the de- tailed structure of the transition region on a highly refined 3-D grid with very short time steps.

#### 3.2.2 Alfvén-Wave-driven SOlar Real Time Model (AWSoM-R)

The critical part of the SC model is the transition region and low corona, which extends from the upper chromosphere to the heliocentric distances of about $(1.03 \div 1.15)R_\odot$. In this thin shell around the Sun, the Alfvén waves pass from the chromosphere to the SC, the plasma temperature increases from ten thousand to millions K, and the solar wind forms. At the same time, this transition region is also a computational bottleneck, which decreases the efficiency and performance of the entire SC model. The temperature gradients across the transition region are very sharp, and the magnetic field topology is complicated at small altitudes above the solar surface. A very fine grid is needed to resolve such gradients and a fully 3-D implicit solver to calculate the field-aligned electron heat conduction.

A simple but convenient way to simulate a magnetically-driven CME is to superimpose a (Gibson & Low, 1998) (GL) or (Titov & D´emoulin, 1999) (TD) magnetic flux-tube configuration onto the background state of the SC. Specific examples of such CME simulations using the AWSoM model for the SC and IH with a superimposed GL magnetic configuration include (Manchester et al., 2012). The GL magnetic configuration describes an erupting magnetic filament filled with excessive plasma density. That filament becomes an expanding flux rope (magnetic cloud) in the ambient solar wind while evolving and propagating outward from the Sun, thus allowing the simulation of the propagation to 1 AU of a magnetically driven CME. Similarly, by superimposing multiple TD configurations, (Linker et al., 2016) have recently modeled the July 2000 CME eruption, which was accompanied by the extreme SEP event.

This degradation of the *3-D model* performance can be avoided by recognizing that the dominant physical processes in the low SC are magnetic field-aligned, and, therefore, may be mathematically, described by a multitude of 1-D equation sets along the field lines. Indeed, the heat fluxes and low velocities in a low beta plasma are aligned with the magnetic field as well as the Alfv´en wave energy flux. Within the computational model, AWSoM-R (ASWoM - Real-time) developed by (Sokolov et al., 2016), a multitude of magnetic field lines – *"threads"* – are traced downward from all grid points of the lower boundary of the AWSoM model describing the SC. This boundary is set at a heliocentric distance of $R = R_b \sim (1.03 \div 1.15) R_\odot$ above the transition region so that above the boundary ($R \geq R_b$) the AWSoM model can be solved on a moderate resolution grid. Meanwhile, below the boundary, at $R_\odot < R < R_b$, one can readily solve the plasma parameters along each thread using 1-D equations, thus connecting each boundary point to the upper chromosphere and providing the boundary condition for the AWSoM model. One still must use a high-resolution grid on the section of the thread within the transition region; however, with modern computers, any resolution requirement for 1-D models is not a problem.

### 3.3 SEP Transport and Acceleration: Focused-Transport Equation

From observations, one can see a pronounced difference between the *gradual* and *flare* (or *impulsive*) SEP events (Reames, 1999; Vainio, 2009). In the course of gradual events, the SEP flux at 1 AU increases slowly and is believed to result from the diffusive shock acceleration (DSA) (Krymskii, 1977; Axford et al., 1977; Blandford & Ostriker, 1978; Bell, 1978). In the vicinity of the shock, the divergence of the plasma flow velocity has a large negative value. Hence, the plasma flow is highly convergent. Due to scattering from turbulent waves, an ion can be effectively trapped in this region between two converging semi-transparent mirrors, one located upstream and the other downstream the shock. Under such circumstances, the particle undergoes *first order* (Fermi, 1954) acceleration. In contrast with the gradual events, flare-accelerated SEPs have very fast raise times (∼1 hr), when the shock can hardly appear. This sort of SEP event is believed to be caused by the particle acceleration during flares, probably, by stochastic acceleration ( or second-order Fermi acceleration). In contrast with the DSA, the flare acceleration is fast and mainly takes place near the Sun.

Mathematically, the most advanced SEP models are based on solving the focused transport equation (Eq. 1) (Skilling, 1971; Isenberg, 1997; Ruffolo, 1995; Kóta & Jokipii, 1997), describing the field-aligned transport of charged particles. The focused-transport approach is suitable to describe anisotropic particle distributions as long as the full pitch-angle distribution is included.

#### *3.3.1 Adaptive Mesh Particle Simulator (AMPS)*

Energetic solar ions and electrons are effectively tied to magnetic field lines, which, in turn, are tightly frozen into the solar wind plasma. One can consider a single field line, which evolves as it is carried by the solar wind. SEPs are advected together with the field line and can only move along their "host" field line. In this approximation, the SEP transport can be conveniently described by the kinetic equation in Lagrangian coordinates. The remarkable property of these coordinates (Landau & Lifshitz, 1959) is that any point on a selected field line with the given Lagrangian coordinate moves in the 3-D space with the local solar wind velocity. Both the particle motion along the field line and the transport equation in Lagrangian coordinates are essentially 1-D problems. Moreover, the transport coefficients in this 1-D equation depend solely on the MHD quantities of the selected field line, as had been demonstrated by Sokolov et al. (2004); Kóta and Jokipii (2004); Kóta et al. (2005)

AMPS was originally developed targeting planetary applications (Tenishev et al., 2021). Later, AMPS became a fully integrated components of the SWMF playing the role in plasma modeling (Shou et al., 2021). Recently, the code was adapted for modeling SEPs ion the inner heliosphere by solving both Parker and Focused transport equations for SEPs moving along a set of evolving magnetic field lines (Tenishev & Combi, 2005).

In simulating SEPs transport and acceleration of SEPs in the heliosphere, AMPS works in concert with other components of the SWMF. Particularly, a coupling is done between SWMF/SC, SWMF/IH, SWMF/M-FLAMPA, and SWMF/AMPS components. Here the first two provides the background plasma and turbulence, M-FLAMPA served a tool for extracting the evolving magnetic field lines, and AMPS, eventually, simulated transport of SEPs as illustrated in Fig. 1.

**References**

Axford, W. I., Leer, E., & Skadron, G. (1977). The Acceleration of Cosmic Rays by Shock Waves. *Proc. 15th Int. Cosmic Ray Conf.*, *11*, 132-137.

Band, D., Matteson, J., Ford, L., Schaefer, B., Palmer, D., Teegarden, B., ... Lestrade, P. (1993). Batse observations of gamma-ray burst spectra. i. spectral diversity. *Astrophysical Journal*, *413*, 281.

Bell, A. R. (1978, January). The Acceleration of Cosmic Rays in Shock Fronts. I. *Monthly Notices of Royal Astron. Soc.*, *182*, 147-156.

Blandford, R. D., & Ostriker, J. P. (1978, April). Particle Acceleration by Astrophysical Shocks. *Astrophys. J. Lett.*, *221*, L29-L32. doi: 10.1086/182658

Borovikov, D., Sokolov, I. V., Huang, Z., Roussev, I. I., & Gombosi, T. I. (2019). Toward quantitative model for simulation and forecast of solar energetic particle production during gradual events – ii: kinetic description of sep. *arXiv:1911.10165*.

Cane, H. V., Reames, D. V., & von Rosenvinge, T. T. (1988). The role of interplanetary shocks in the longitude distribution of solar energetic particles. *Journal of Geophysical Research*, *93*(A9), 9555.

Cane, H. V., von Rosenvinge, T. T., Cohen, C. M. S., & Mewaldt, R. A. (2003). Two components in major solar particle events. *Geophysical Research Letters*, *30*(12), 8017.

Cliver, E. W. (2006). The american astronomical society, find out more the institute of physics, find out more the unusual relativistic solar proton events of 1979 august 21 and 1981 may 10. *Astrophysical Journal*, *639*(2), 1206.

Cohen, C. M. S., Mewaldt, R. A., Cummings, A. C., Leske, R. A., Stone, E. C., von Rosenvinge, T. T., & Wiedenbeck, M. E. (2003). Variability of spectra in large solar energetic particle events. *Advances in Space Research*, *32*(12), 2649-2654.

Cohen, O., Sokolov, I. V., Roussev, I. I., Arge, C. N., Manchester, W. B., Gombosi, T. I., ... Velli, M. (2007, January). A Semiempirical Magnetohydrodynamical Model of the Solar Wind. *Astrophysical Journal*, *654*, L163-L166. doi: 10.1086/511154

Desai, M. I., Mason, G. M., Dwyer, J. R., Mazur, J. E., Gold, R. E., Krimigis, S. M., ... Skoug, R. M. (2003, may). Evidence for a suprathermal seed population of heavy ions accelerated by interplanetary shocks near 1 AU. *The Astrophysical Journal*, *588*(2), 1149–1162. Retrieved from https://doi.org/10.1086/374310 doi: 10.1086/374310

Desai, M. I., Mason, G. M., Gold, R. E., Krimigis, S. M., Cohen, C. M. S., Mewaldt, R. A., ... Dwyer, J. R. (2006, sep). Heavy-ion elemental abundances in large solar energetic particle events and their implications for the seed population. *The Astrophysical Journal*, *649*(1), 470–489. Retrieved from https://doi.org/10.1086/505649 doi: 10.1086/505649

Desai, M. I., Mason, G. M., Mazur, J. E., & Dwyer, J. R. (2006). The seed population for energetic particles accelerated by cme-driven shocks. *Space Science Reviews*, *124*, 261–275.

Desai, M. I., Mason, G. M., Wiedenbeck, M. E., Cohen, C. M. S., Mazur, J. E., Dwyer, J. R., ... Skoug, R. M. (2004, aug). Spectral properties of heavy ions associated with the passage of interplanetary shocks at 1 AU. *The Astrophysical Journal*, *611*(2), 1156–1174. Retrieved from https://doi.org/10.1086/422211 doi: 10.1086/422211

Dresing, N., Gómez-Herrero, R., Klassen, A., Heber, B., Kartavykh, Y., & Dröge, W. (2012). The large longitudinal spread of solar energetic particles during the 17 january 2010 solar event. *Solar Physics*, *281*, 281–300.

Dröge, W., Kartavykh, Y. Y., Klecker, B., & Kovaltsov, G. A. (2010). Anisotropic three-dimensional focused transport of solar energetic particles in the inner heliosphere. *Astrophysical Journal*, *709*, 912–919,.

Dröge, W., Kartavykh, Y. Y., Klecker, B., & Mason, G. M. (2006). Acceleration and transport modeling of solar energetic particle charge states for the event of 1998 september 9. *Astrophysical Journal*, *645*, 1516–1524.

Evans, R. M., Opher, M., & Gombosi, T. I. (2011, February). Learning from the outer heliosphere: Interplanetary coronal mass ejection sheath flows and the ejecta orientation in the lower corona. *Astrophys. J.*, *728*, 41. doi: 10.1088/0004-637X/728/1/41

Fermi, E. (1954, January). Galactic magnetic fields and the origin of cosmic radiation. , *119*, 1. doi: 10.1086/145789

Gibson, S. E., & Low, B. C. (1998, January). A Time-Dependent Three-Dimensional Magnetohydrodynamic Model of the Coronal Mass Ejection. *Astrophys. J.*, *493*, 460-473.

Groth, C. P. T., DeZeeuw, D. L., Gombosi, T. I., & Powell, K. G. (2000, November). Global three-dimensional mhd simulation of a space weather event: CME formation, interplanetary propagation, and interaction with the magnetosphere. *J. Geophys. Res.*, *105*, 25,053—25,078.

He, H.-Q., Qin, G., & Zhang, M. (2011, may). PROPAGATION OF SOLAR ENERGETIC PARTICLES IN THREE-DIMENSIONAL INTERPLANETARY MAGNETIC FIELDS: IN VIEW OF CHARACTERISTICS OF SOURCES. *The Astrophysical Journal*, *734*(2), 74.

He, H.-Q., & Wan, W. (2015). Numerical study of the longitudinally asymmetric distribution of solar energetic particles in the heliosphere. *Astrophysical Journal*, *218*, 17.

He, H.-Q., & Wan, W. (2019). Propagation of solar energetic particles in the outer heliosphere: Interplay between scattering and adiabatic focusing. *The Astrophysical Journal Letters*, *885* (L28).

Hellweg, C. E., & Baumstark-Khan, C. (2007, jan). Getting ready for the manned mission to Mars: The astronauts' risk from space radiation. *Naturwissenschaften*, *94*(7), 517–526. doi: 10.1007/s00114-006-0204-0

Ho, T. M., Thomas, N., Boice, D. C., Kollein, C., & Soderblom, L. A. (2003). Comparative study of the dust emission of 19p/borrelly (deep space 1) and 1p/halley. *Advances in Space Research*, *31*(12), 2583-2589.

Hu, J., Li, G., Ao, X., Zank, G. P., & Verkhoglyadova, O. (2017). Modeling particle acceleration and transport at a 2-d cme-driven shock. *Journal of Geophysical Research*, *122*, 10,938–10,963.

Isenberg, P. A. (1997, March). A Hemispherical Model of Anisotropic Interstellar Pickup Ions. *J. Geophys. Res.*, *102*, 4,719-4,724.

Jackman, C. H., DeLand, M. T., Labow, G. J., Fleming, E. L., Weisenstein, D. K., Ko, M. K. W., ... Russell, J. M. (2005). Neutral atmospheric influences of the solar proton events in october–november 2003. *Journal of Geophysical Research*, *110*, A09S27.

Jacobs, C., Roussev, I. I., Lugaz, N., & Poedts, S. (2009, April). The Internal Structure of Coronal Mass Ejections: Are all Regular Magnetic Clouds Flux Ropes? *Astrophys. J. Lett.*, *695*, L171-L175. doi: 10.1088/0004-637X/695/2/L171

Jäkel, O. (2004). Radiation hazard during a manned mission to Mars. *Zeitschrift fur medizinische Physik*, *14*(4), 267–272.

Jin, M., Manchester, W. B., van der Holst, B., Oran, R., Sokolov, I., Tóth, G., ... Gombosi, T. I. (2013). Numerical simulations of coronal mass ejection on 2011 March 7: One-temperature and two-temperature model comparison. , *773*(1), 50. doi: 10.1088/0004-637X/773/1/50

Jin, M., Manchester, W. B., van der Holst, B., Sokolov, I., Tóth, G., Mullinix, R. E., ... Gombosi, T. I. (2017). Data-constrained coronal mass ejections in a global magnetohydrodynamics model. *The Astrophysical Journal*, *834*(2), 173.

Jin, M., Manchester, W. B., van der Holst, B., Sokolov, I., Tóth, G., Vourlidas, A., ... Gombosi, T. I. (2017). Chromosphere to 1 au simulation of the 2011 march 7th event: A comprehensive study of coronal mass ejection propagation. *The Astrophysical Journal*, *834* (2), 172.

Jokipii, J. R. (1966). Cosmic-ray propagation. i. charged particles in a random magnetic field. *Astrophysical Journal*, *146*, 480.

Kóta, J., & Jokipii, J. R. (1997). Energy changes of particles moving along field line. In *Proc. 25th international cosmic ray conference* (Vol. 1, pp. 213–216).

Kóta, J., & Jokipii, J. R. (2004). Cosmic ray acceleration and transport around the termination shock. *AIP Conference Proceedings*, *719* (1), 272-278. doi: http://dx.doi.org/10.1063/1.1809528

Kota, J., Manchester, W. B., Jokipii, J. R., de Zeeuw, D. L., & Gombosi, T. I. (2005). Simulation of sep acceleration and transport at cme-driven shocks. *AIP Conference Proceedings*, *781*, 201-206.

Kóta, J., Manchester, W. B., Jokipii, J. R., de Zeeuw, D. L., & Gombosi, T. I. (2005). Simulation of sep acceleration and transport at cme???driven shocks. *AIP Conference Proceedings*, *781*, 201-206.

Krymskii, G. F. (1977, June). A Regular Mechanism for Accelerating Charged Particles at the Shock Front. *Akademiia Nauk SSSR Doklady*, *234*, 1,306-1,308.

Laitinen, T., Dalla, S., & Marsh, M. (2013). Energetic particle cross-field propagation early in a solar event. *Astrophysical Journal*, *773*, L29.

Landau, L. D., & Lifshitz, E. M. (1959). *Fluid Mechanics*. Pergamon Press: Oxford.

Lario, D., Aran, A., Gomez-Herrero, R., Dresing, N., Heber, B., Ho, G. C., ... Roelof, E. C. (2013). Longitudinal and radial dependence of solar energetic particle peak intensities: Stereo, ace, soho, goes, and messenger observations. *Astrophysical Journal*, *767*, 41.

Lario, D., & Decker, R. B. (2011). Estimation of solar energetic proton mission-integrated fluences and peak intensities for missions traveling close to the sun. *Space Weather*, *9*, S11003.

le Roux, J. A., & Webb, G. M. (2009). Time-dependent acceleration of interstellar pickup ions at the heliospheric termination shock using a focused transport approach. *Astrophysical Journal*, *693*, 534.

Li, G., Shalchi, A., Ao, X., Zank, G., & Verkhoglyadova, O. (2012). Particle acceleration and transport at an oblique cme-driven shock. *Advances in Space Research*, *49*, 1067–1075.

Li, G., & Zank, G. P. (2005). Mixed particle acceleration at cme-driven shocks and flares. *Geophysical Research Letters*, *32*, L02101.

Li, G., Zank, G. P., & Rice, W. K. M. (2003). Energetic particle acceleration and transport at coronal mass ejection–driven shocks. *Journal of Geophysical Research*, *108* (A2), 1082.

Linker, J., Torok, T., Downs, C., Lionello, R., Titov, V., Caplan, R. M., ... Riley, P. (2016). Mhd simulation of the bastille day event. *AIP Conference Proceedings*, *1720* (1). doi: http://dx.doi.org/10.1063/1.4943803

Manchester, W. B., Gombosi, T. I., Roussev, I., De Zeeuw, D. L., Sokolov, I. V., Powell, K. G., ... Opher, M. (2004, January). Three-dimensional MHD simulation of a flux rope driven CME. *J. Geophys. Res.*, *109(A18)*, 1,102-1,119.

Manchester, W. B., Gombosi, T. I., Roussev, I., Ridley, A., De Zeeuw, D. L., Sokolov, I. V., ... Tóth, G. (2004, February). Modeling a space weather event from the Sun to the Earth: CME generation and interplanetary propagation. *J. Geophys. Res.*, *109(A18)*, 2,107–2,122.

Manchester, W. B., van der Holst, B., Toth, G., & Gombosi, T. I. (2012, SEP 1). The coupled evolution of electrons and ions in coronal mass ejection-driven shocks. , *756* (1). doi: 10.1088/0004-637X/756/1/81

Mckibben, R. B. (1972). Azimuthal propagation of low-energy solar-flare protons as observed from spacecraft very widely separated in solar azimuth. *Journal of Geophysical Research*, *77* (22), 3957.

McKibben, R. B., Connell, J. J., Lopate, C., Zhang, M., Anglin, J. D., Balogh, A.,

... Heber, B. (2003). Ulysses cospin observations of cosmic rays and solar energetic particles from the south pole to the north pole of the sun during solar maximum. *Annales Geophysicae*, *21*(6), 1217.

Mewaldt, R. A. (2006). Solar energetic particle composition, energy spectra, and space weather. *Space Science Reviews*, *124*(1), 303–316.

Morris, D. (2007). *From the flight deck: Plane talk and sky science*. Toronto, Ontario, Canada: ECW Press.

Ng, C. K., & Reames, D. V. (2003). Modeling shock-accelerated solar energetic particles coupled to interplanetary alfv́en waves. *The Astrophysical Journal*, *591*(1), 461-485.

Powell, K. G., Roe, P. L., Linde, T. J., Gombosi, T. I., & De Zeeuw, D. L. (1999, September). A solution-adaptive upwind scheme for ideal magnetohydrodynamics. *J. Comp. Phys.*, *154*, 284-309.

Qin, G., He, H.-Q., , & Zhang, M. (2011). An effect of perpendicular diffusion on the anisotropy of solar energetic particles from unconnected sources. *Astrophysical Journal*, *738*(28).

Qin, G., & Shalchi, A. (2009). Pitch-angle diffusion coefficients of charged particles from computer simulations. *Astrophysical Journal*, *707*, 61.

Qin, G., & Wang, Y. (2015). Simulations of a gradual solar energetic particle event observed by helios 1, helios 2, and imp 8. *Astrophysical Journal*, *809*, 177.

Qin, G., Wang, Y., Zhang, M., & Dalla, S. (2013, mar). TRANSPORT OF SOLAR ENERGETIC PARTICLES ACCELERATED BY ICME SHOCKS: REPRODUCING THE RESERVOIR PHENOMENON. *The Astrophysical Journal*, *766*(2), 74. Retrieved from https://doi.org/10.1088/0004-637x/766/2/74 doi: 10.1088/0004-637x/766/2/74

Qin, G., Zhang, M., & Dwyer, J. R. (2006). Effect of adiabatic cooling on the fitted parallel mean free path of solar energetic particles. *Journal of Geophysical Research (Space Physics)*, *111*, A08101.

Reames, D. (1999). Particle acceleration at the sun and in the heliosphere. *Space Science Reviews*, *90*, 413–491.

Reames, D. V. (1999, February). Particle Acceleration at the Sun and in the Heliosphere. *Space Sci. Rev.*, *90*, 413-491.

Reames, D. V., Ng, C. K., & Tylka, A. J. (2013). Spatial distribution of solar energetic particles in the inner heliosphere. *Solar Physics*, *285*, 233–250.

Roelof, E. C., Gold, R. E., Simnett, G. M., Tappin, S. J., Armstrong, T. P., & Lanzerotti, L. J. (1992). Low-energy solar electrons and ions observed at ulysses february-april, 1991: The inner heliosphere as a particle reservoir. *Geophysical Research Letters*, *19*(12), 1243.

Roussev, I. I., Forbes, T. G., Gombosi, T. I., Sokolov, I. V., DeZeeuw, D. L., & Birn, J. (2003, May). A three-dimensional flux rope model for coronal mass ejections based on a loss of equilibrium. *Astrophys. J. Lett.*, *588*, L45-L48.

Roussev, I. I., Lugaz, N., & Sokolov, I. V. (2007, October). New physical insight on the changes in magnetic topology during coronal mass ejections: Case studies for the 2002 April 21 and August 24 events. *Astrophys. J. Lett.*, *668*, L87–L90. doi: 10.1086/522588

Roussev, I. I., Sokolov, I. V., Forbes, T. G., Gombosi, T. I., Lee, M. A., & Sakai, J. I. (2004, April). A numerical model of a coronal mass ejection: Shock development with implications for the acceleration of gev protons. *Astrophys. J. Lett.*, *605*, L73–L76.

Ruffolo, D. (1995). Effect of adiabatic deceleration on the focused transport of solar cosmic rays. *Astrophysical Journal*, *442*, 861-874.

Ruffolo, D. (1995, April). Effect of adiabatic deceleration on the focused transport of solar cosmic rays. , *442*, 861-874. doi: 10.1086/175489

Schwadron, N. A., Townsend, L., Kozarev, K., Dayeh, M. A., Cucinotta, F., Desai, M., ... Squier, R. K. (2010). Earth-moon-mars radiation environment module framework. *Space Weather*, *8*(S00E02).

Shou, Y., Tenishev, V., Chen, Y., Toth, G., & Ganushkina, N. (2021). Magnetohydrodynamic with adaptively embedded particle-in-cell model: Mhd-aepic. *Journal of Computational Physics*, *446*, 110656. Retrieved from https://www.sciencedirect.com/science/article/pii/S0021999121005519 doi: https://doi.org/10.1016/j.jcp.2021.110656

Skilling, J. (1971). Cosmic rays in the galaxy: Convection or diffusion? *Astrophysical Journal*, *170*(265-273), 265-273.

Skilling, J. (1971, December). Cosmic Rays in the Galaxy: Convection or Diffusion? *Astrophys. J.*, *170*, 265-273.

Sokolov, I. V., B.van der Holst, Manchester, W. B., Ozturk, D., Szente, J., Taktakishvili, A., . . . Gombosi, T. I. (2016). Threaded-field-lines model for the low solar corona powered by Alfvén wave turbulence. *arXiv:1609.04379*, *[astro-ph.SR]*.

Sokolov, I. V., Roussev, I. I., Gombosi, T. I., Lee, M. A., Kóta, J., Forbes, T. G., . . . Sakai, J. I. (2004, December). A new field-line-advection model for solar particle acceleration. *Astrophys. J. Lett.*, *616*, L171–L174.

Sokolov, I. V., Roussev, I. I., Skender, M., Gombosi, T. I., & Usmanov, A. V. (2009, May). Transport Equation for MHD Turbulence: Application to Particle Acceleration at Interplanetary Shocks. *Astrophys. J.*, *696*, 261-267. doi: 10.1088/0004-637X/696/1/261

Tan, L. C., Reames, D. V., Ng, C. K., Saloniemi, O., & Wang, L. (2009). Observational evidence on the presence of an outer reflecting boundary in solar energetic particle events. *Astrophysical Journal*, *701*(2), 1753-1764.

Tenishev, V., & Combi, M. R. (2005). Monte-Carlo model for dust/gas interaction in rarefied flows. *AIAA-2005-4832*.

Tenishev, V., Shou, Y., Borovikov, D., Lee, Y., Fougere, N., Michael, A., & Combi, M. R. (2021). Application of the Monte Carlo method in modeling dusty gas, dust in plasma, and energetic ions in planetary, magnetospheric, and heliospheric environments. *Journal of Geophysical Research: Space Physics*, *126*, e2020JA028242. doi: https://doi.org/10.1029/2020JA028242

Titov, V. S., & D´emoulin, P. (1999, November). Basic Topology of Twisted Magnetic Configurations in Solar Flares. *Astron. & Astrophys.*, *351*, 707-720.

Tylka, A. J. (2001). New insights on solar energetic particles from wind and ace. *Journal of Geophysical Research (Space Physics)*, *106*(A11), 25,333.

Tylka, A. J., Cohen, C. M. S., Dietrich, W. F., Lee, M. A., Maclennan, C. G., Mewaldt, R. A., . . . Reames, D. V. (2006). A comparative study of ion characteristics in the large gradual solar energetic particle events of 2002 april 21 and 2002 august 24. *Astrophysical Journal Supplement Series*, *164*(2), 536-551.

Vainio, R. (2009, March). Particle acceleration and turbulence transport in heliospheric plasmas. In N. Gopalswamy & D. F. Webb (Eds.), *Universal heliophysical processes* (Vol. 257, p. 413-423). doi: 10.1017/S1743921309029640

Wang, Y., & Qin, G. (2004). Estimation of the release time of solar energetic particles near the sun. *arXiv:1311.7469v4*.

Young, M. A., Schwadron, N. A., Gorby, M., Linker, J., Caplan, R. M., Downs, C., . . . Cohen, C. M. S. (2021). Energetic proton propagation and acceleration simulated for the bastille day event of 2000 july 14. *Astrophysical Journal*, *909*(160).

Zank, G. P., Li, G., Florinski, V., Matthaeus, W. H., Webb, G. M., & le Roux, J. A. (2004). Perpendicular diffusion coefficient for charged particles of arbitrary energy. *Journal of Geophysical Research*, *109*, A04107.

Zhang, M., Qin, G., & Rassoul, H. (2009). PROPAGATION OF SOLAR ENERGETIC PARTICLES IN THREE-DIMENSIONAL INTERPLANETARY MAGNETIC FIELDS. *The Astrophysical Journal*, *692*(1), 109–132.

Zhao, L., & Li, G. (2014). Particle acceleration at a pair of parallel shocks near the sun. *Journal of Geophysical Research (Space Physics)*, *119*, 6106–6119.